\documentclass[12pt]{article}
\usepackage{epsfig}
\begin{document}
\begin{center}
{\bf \large Mechanism of Pion Production in $\alpha p$
Scattering at 1~GeV/nucleon}\\
\vspace{5mm}
 G. D.~Alkhazov, A. V.~Kravtsov \footnote{deceased},
A. N.~Prokofiev, I. B.~Smirnov\\
 (the SPES4-$\pi$ Collaboration)\\
\vspace{5mm}
 {\it St.Petersburg Nuclear Physics Institute,
 Gatchina, RUSSIA }\\
\end{center}
\vspace{10mm}
\par
The one-pion and two-pion production in the $p(\alpha ,\alpha
^\prime) X$ reaction at an energy of $E_\alpha$~=~4.2~GeV has been
studied by simultaneous registration of the scattered $\alpha
^{\prime}$ particle and the secondary proton or pion. The obtained
results demonstrate that the inelastic $\alpha$-particle
scattering on the proton at the energy of the experiment proceeds
either through excitation and decay of the $\Delta$ resonance in
the projectile $\alpha$ particle, or through excitation  in the
target proton of the Roper resonance, which decays mainly on a
nucleon and a pion or a nucleon and a $\sigma$ meson
-- a system of two pions in the isospin $I$~=~0,~$S$-wave state.\\
\par
\noindent {\bf Comments:} 16 pages, 10 figures. Submitted to
Proceedings of the XX International Baldin Seminar on High-Energy
Physics Problems, Dubna, October 4-9, 2010.\\
\par
\noindent {\bf Report:} St.Petersburg Nuclear Physics Institute
preprint
PNPI-2009, 2791 (2009).\\
\par
\noindent {\bf Category:} Nuclear Experiment (nucl-ex); High-Energy Physics Experiment (hep-ex).\\

\clearpage
\noindent {\bf 1 Introduction}\\

\noindent A study of inelastic $\alpha$$p$ scattering at an energy
of $\sim$ 1~GeV/nucleon is of significant interest since it is
related, in particular, to the problem of the
$N(1440)P_{11}$~(Roper) resonance. The Roper resonance [1] is the
lowest positive-parity excited state $N^*$ of the nucleon, and in
many respects it is a very intriguing and important resonance.
Morsch {\it et al.} [2, 3] have interpreted the excitation of the
Roper resonance in inelastic $\alpha p$ scattering as the
breathing-mode ($L$~=~0) monopole excitation of the nucleon. In
this interpretation, the $N(1440)$ resonance mass is related to
the compressibility of the nuclear matter (on the nucleonic
level). This resonance also plays an important role in many
intermediate-energy processes, in the three-body nuclear forces
and in the swelling of nucleons in nuclei. The investigation of
the $N(1440)$ resonance was the goal of numerous theoretical and
experimental studies. This activity was motivated by the still not
properly understood nature of the resonance, its relatively low
mass and anomalously large width of a few hundred MeV.

The Roper resonance was observed and studied for the first time in
the $\pi$$p$ scattering partial-wave analyses [1, 4--7]. The fact
that the Roper resonance is also strongly excited in $\alpha p$
scattering was quite puzzling. To understand the excitation of
this resonance in different reactions, Morsch and Zupranski [3]
performed a combined analysis of the data of $\pi N$-, $\alpha p$-
and $\gamma p$-scattering experiments, with the conclusion that
the $N(1440)$ state represents a structure formed of two
resonances, one understood as the nucleon breathing mode and the
other one as an excited state of the $\Delta_{3,3}(1232)$
($\Delta$) resonance. The first resonance is strongly excited by
scalar probes, like in $\alpha p$ scattering, whereas the second
one is excited in spin-isospin-flip reactions, like in $\pi N$
scattering.  The two-resonance picture of $N(1440)$ and the
breathing-mode excitation of the proton was also discussed by the
same authors [8] in a reanalysis of high-energy $pp$- and $\pi
p$-scattering data.

An advantage of studying the Roper resonance in an $\alpha
p$-scattering experiment, as compared to $\pi N$, $NN$ and $\gamma
N$ experiments, is that in the case of $\alpha p$ scattering
 the number of the reaction channels is
rather limited. At an energy of $\sim$ 1 GeV/nucleon, the Roper
resonance is strongly excited in $\alpha p$ scattering, whereas
the contribution from excitation of other baryon resonances is
expected to be small [9].

Inelastic $\alpha p$ scattering was investigated previously at
$E_\alpha$ = 4.2~GeV in an inclusive experiment [2] at the
Saturne-II accelerator in Saclay using the SPES4 magnetic
spectrometer. The energy distribution  of the scattered $\alpha$
particles from the $p(\alpha ,\alpha^{\prime})X$ reaction was
studied, and a strong excitation of the $N(1440)$ state was found.
Two peaks were observed in the missing-energy,
$\omega$~=~$E_{\alpha ^\prime}$ -- $E_\alpha$, distribution
(Fig.~1). A large one, in the region of small energy transfers,
$\omega\simeq$~--~0.25~GeV, was evidently due to excitation of the
$\Delta$ resonance in the projectile $\alpha$ particle, and a
smaller one, in the region of $\omega\simeq$~--~0.55~GeV, was
interpreted by Morsch {\it et al.} [2] as a signal of the Roper
resonance excitation in the target proton. This interpretation was
confirmed later by a more detailed theoretical consideration of
Hirenzaki $et~al.$ [10, 11].

\begin{figure}[htb]
\centering\epsfig{file=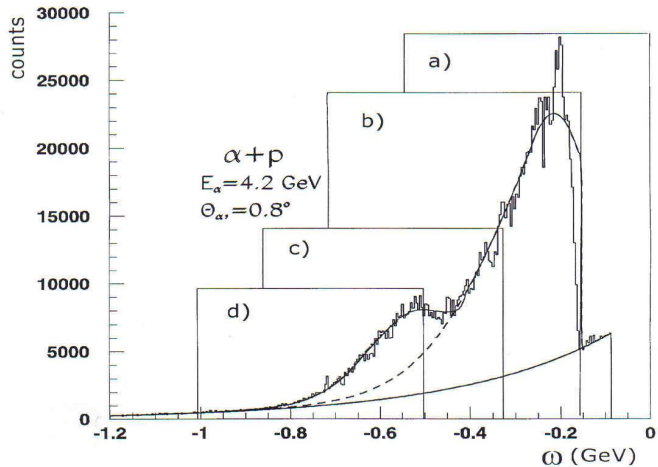,width=0.65 \textwidth,height=65mm
}\caption{\small Inclusive missing-energy ($\omega$) spectrum of
inelastic $\alpha p$ scattering [2]. The acceptance boundaries of
$\omega$ for different SPES4 momentum settings in the present
 experiment are marked as (a), (b), (c) and (d). The mean
values of these intervals correspond to $q_{\alpha^\prime}/Z$ =
3.35; 3.25; 3.15 and 3.06~GeV/$c$, respectively.}
\end{figure}

According to theory [10], only three reaction channels dominate in
inelastic $\alpha p$ scattering at this energy.  The first one
(Fig. 2a) corresponds to excitation of the $\Delta$ resonance in
the $\alpha$-particle projectile, while the second and third ones
(Figs. 2b,c) correspond to excitation of the Roper (or
$N$(1520)$D_{13}$) resonance in the target proton mainly through
exchange of a neutral ``sigma meson" ($\sigma$). The contribution
of other channels is practically negligible. Note that due to the
isoscalar nature of the $\alpha$ particle and isospin
conservation, direct excitation of the $\Delta$ resonance in the
proton is forbidden. The final-state products from the $p(\alpha
,\alpha^{\prime})X$ reaction may be either a nucleon (proton or
neutron) and one pion, resulting from decay of the $\Delta$ or
Roper resonances (Fig. 2, diagrams a, b), or a nucleon and two
pions, resulting from decay of the Roper resonance \mbox{(Fig. 2,
diagram c)}.
\begin{figure}
\centering\epsfig{file=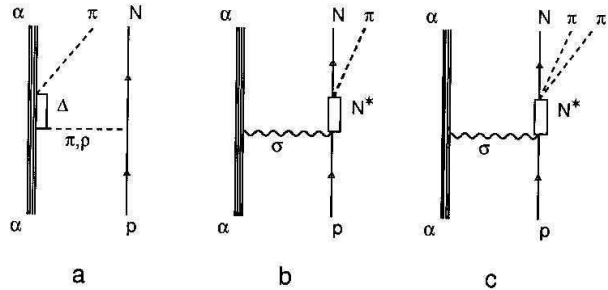,width=0.75\textwidth,height=70mm}\caption{\small
Main diagrams contributing to the $p(\alpha, \alpha^{\prime})X$
reaction: (a) $\Delta$ excitation in the projectile, (b) $N^*$
excitation in the target with the following one-pion ($N\pi$)
decay, and (c) $N^*$ excitation in the target with the following
two-pion ($N\pi\pi$) decay.}
\end{figure}

A drawback of the inclusive $\alpha p$ experiment [2] was that
only the momentum of the scattered $\alpha$ particles was
measured, while other reaction products were not detected. In
order to get more information on the mechanism of the $\alpha
p$-scattering reaction, a semi-exclusive experiment at the
Saturne-II accelerator (Saclay) was performed [12], in which the
decay products as well as the scattered $\alpha$ were registered.

Here we discuss results of this experiment. The conclusions
drawn in this work are based on the missing mass Dalitz plots
analysis and on comparisons of the shapes of the experimental
spectra with those of the simulated ones.\\
\newpage
\noindent
{\bf 2 Experiment}\\

\noindent The experimental study of the $p(\alpha
,\alpha^\prime)X$ reaction discussed in the present paper was
carried out at the Saturne-II accelerator beam of $\alpha$
particles with a momentum $q_{\alpha}$ =~7 GeV/$c$
($E_{\alpha}$~=~4.2~GeV). The scattered $\alpha$ projectiles and
the charged products ($p$, $\pi^+$ or $\pi^-$ ) of the reaction
were registered with the SPES4-$\pi$ set-up [13]. The SPES4-$\pi$
installation included the high-resolution magnetic spectrometer
SPES4, which was also used in earlier experiments, and a
wide-aperture non-focusing Forward Spectrometer (FS). The last one
consisted of an analyzing large-gap dipole magnet, a drift-chamber
telescope and a hodoscope of scintillation counters. A
liquid-hydrogen target, 60~mm in length, was located inside the
analyzing magnet.

The $\alpha$ particles scattered on the liquid-hydrogen target at
an angle of 0.8$^\circ\pm 1.0^\circ$ were registered with SPES4.
The experiment was carried out at four magnetic-rigidity settings
of the SPES4 spectrometer. The central values of
$q_{\alpha^\prime}/Z$ = 3.35, 3.25, 3.15, and 3.06 GeV/$c$ (where
$q_{\alpha^\prime}$ is the momentum of the scattered $\alpha$
particle, and $Z$ = 2 is the $\alpha$-particle charge) were
chosen, which gave us an opportunity to study the reaction at the
energy transfer $\omega$ from -- 0.15 to -- 0.9~GeV. The $\omega$
intervals accepted at different momentum settings of SPES4 in
comparison with the results of the inclusive experiment [2] are
indicated in Fig. 1. The measurements were performed with the full
as well as empty targets. These measurements, properly normalized
to the monitor counts, were used to subtract the background from
the beam halo and from the beam interaction with the target
housing.

The FS allowed to identify the secondary charged particles ($p$,
$\pi^+$, or $\pi^-$) and to reconstruct their trajectories and
momenta. The identification of the particles in the FS was
performed on the basis of the energy-loss and time-of-flight
measurements [13] by means of the scintillator-counter hodoscope.
In the present paper, the data obtained by detecting the scattered
$\alpha$ particles with SPES4 and protons or $\pi ^+$ with the FS
are discussed. The measured momenta $\vec{q}$ of the scattered
$\alpha$ particle and secondary particles were used to determine
the missing mass $M_{\rm miss}$ and the invariant masses
$M$($N\pi$) and $M$($\alpha^{\prime} \pi$) for the one-pion
production channel, and $M$($N\pi\pi$) and $M$($\alpha^{\prime}
\pi\pi$) for the two-pion production channel. In the case when
protons are detected with the FS, the missing mass $M_{\rm miss}$
is the mass of the object $X$ in the $p(\alpha,\alpha^\prime)pX$
reaction, the object $X$ consisting of one or two pions. In the
case when $\pi^+$-pions are detected with the FS, the missing mass
$M_{\rm miss}$ is the mass of the object $X$ in the
$p(\alpha,\alpha^\prime)\pi^+ X$ reaction, the object $X$
consisting of a neutron or a neutron and a $\pi^0$-pion. It should
be noted that the SPES4-$\pi$ set-up has a rather high acceptance
for registration of events from decay of the Roper resonance
(mainly at the momentum settings 3.15 and 3.06~GeV/c), the latter
having the Breit-Wigner (BW) resonance mass at about 1440
MeV/$c^2$ [14]. The SPES4-$\pi$ set-up and the method of the
tracks reconstruction are described in detail in [13].\\

\noindent
{\bf 3 Results and discussion}\\
\noindent {\bf 3.1 Missing mass spectra}\\

\noindent The left panel of Fig. 3 presents the distributions of
events as a function of the missing mass  $M_{\rm miss}$ for the
four momentum settings of SPES4 in the cases of $\pi^+$
registration by the FS. The spectra include the sums of events
from the one-pion and two-pion production channels. For the
momentum setting $q_{\alpha^\prime}$/Z = 3.35 GeV/$c$, only one
peak at $M_{\rm miss}~\simeq M_n$ = 0.94 GeV/$c^2$ corresponding
to the mass of the neutron is seen. The width of this peak
reflects the experimental resolution of the reconstructed values
of $M_{\rm miss}$. Evidently, this peak contains one-pion events
appearing due to decay of $\Delta$ excited in the projectile
$\alpha$ particle. For the momentum settings $q_{\alpha^\prime}$/Z
= 3.15 and 3.06 Gev/$c$, contributions of two-pion events (at
$M_{\rm miss}
> M_n + M_{\pi}$) are observed.

\begin{figure}[htb]
\includegraphics[width=0.47\textwidth,height=100mm]{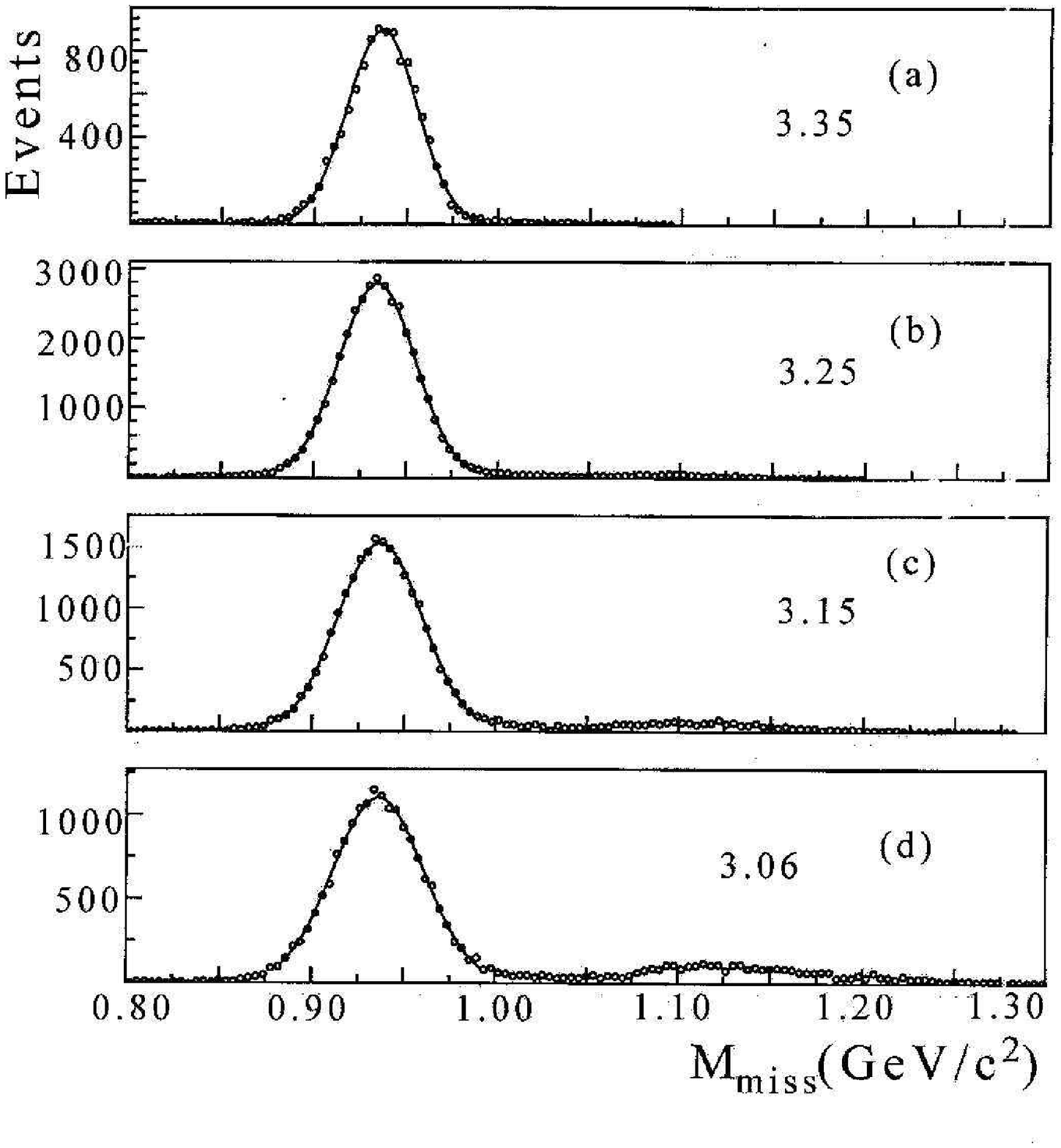}
\includegraphics[width=0.47\textwidth,height=100mm]{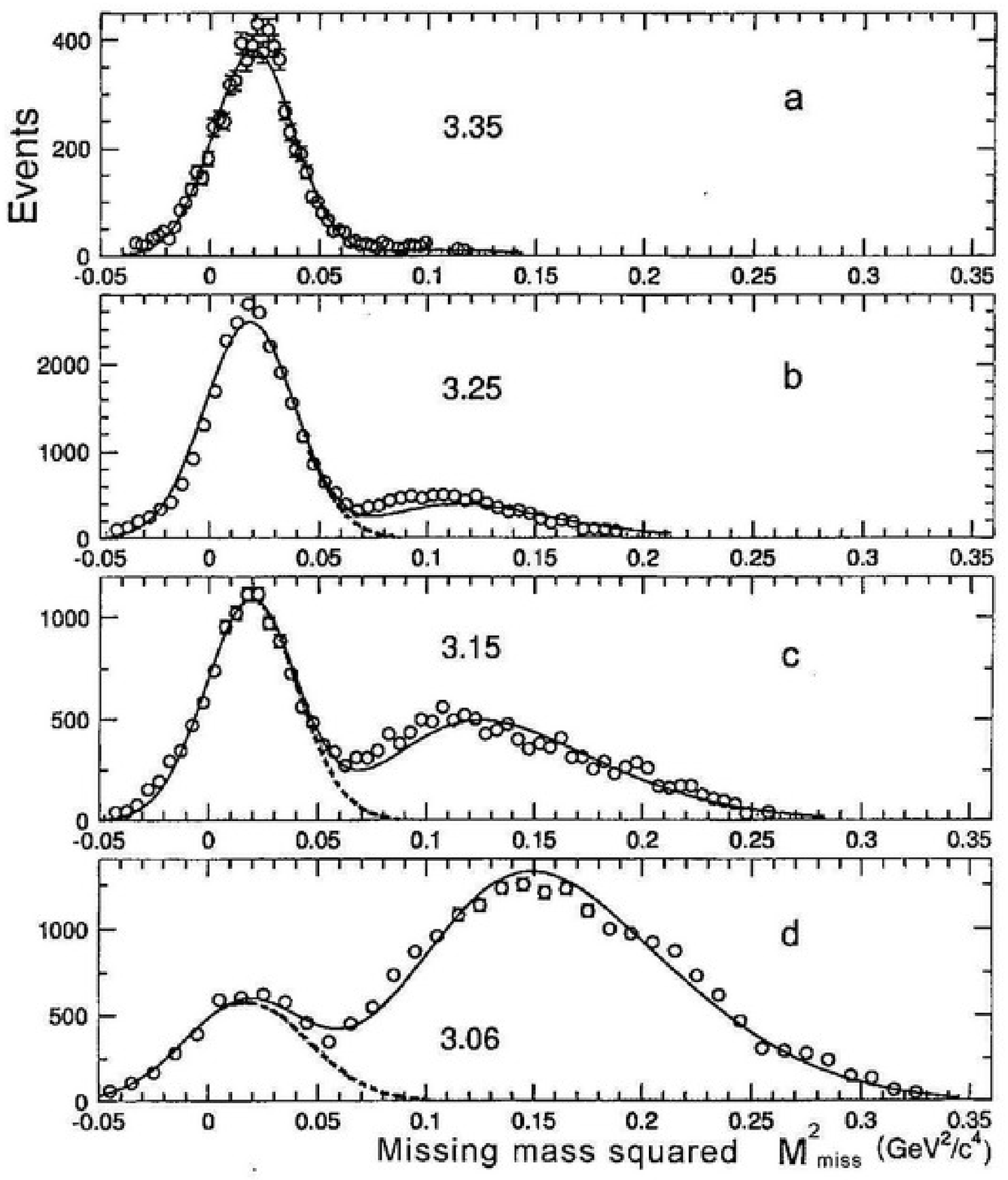}\\
\\
\parbox[t]{0.95\textwidth}{\caption{\small Missing mass spectra for the $p(\alpha,
\alpha^{\prime})\pi^+X$ (left panel) and $p(\alpha,
\alpha^{\prime})pX$ (right panel) reactions for the SPES4 momentum
settings $q_{\alpha^{\prime}}/Z$ = 3.35 (a), 3.25 (b), 3.15 (c),
and 3.06 (d) GeV/$c$. Dots are experimental points. The
distributions of one-pion events are described by Gaussians (solid
lines in the left panel, dashed lines in the right panel). The
solid lines in the right panel are the sums of the
one-pion-production distributions and the simulated
two-pion-production distributions normalized to the experimental
data.}}
\end{figure}

\begin{figure}[htb]
\includegraphics[width=0.31\textwidth,height=55mm]{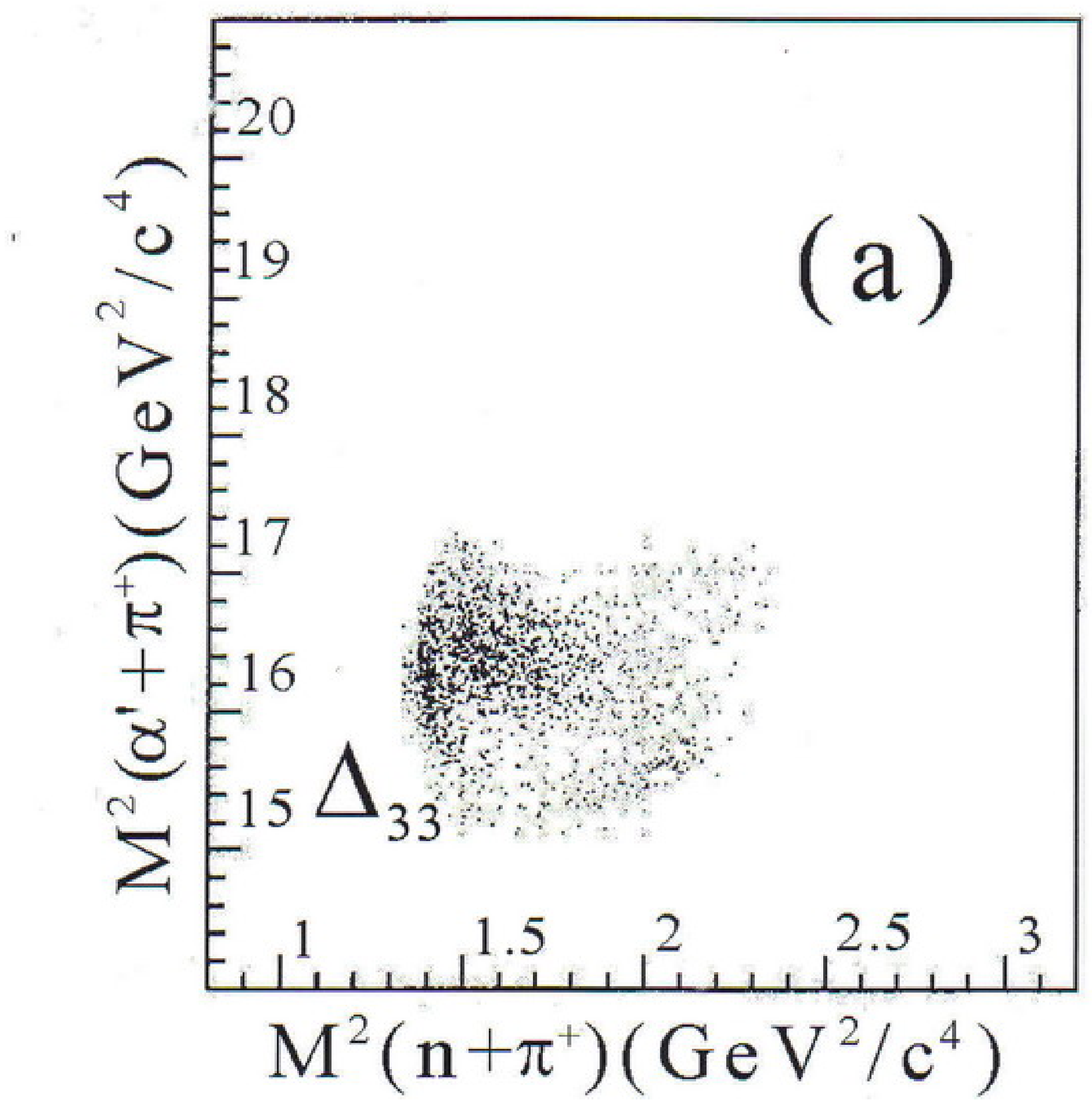}
\includegraphics[width=0.31\textwidth,height=55mm]{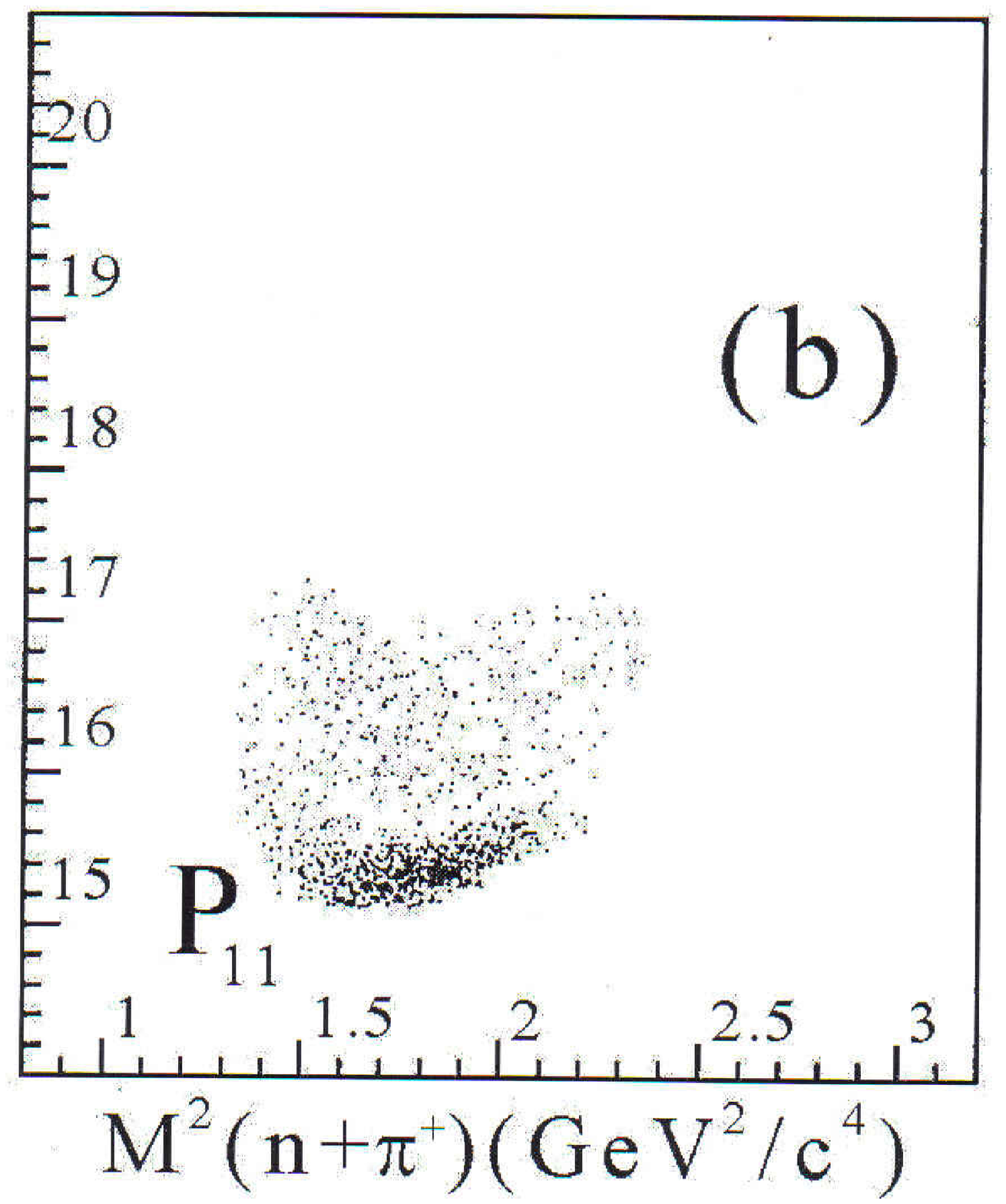}
\includegraphics[width=0.31\textwidth,height=55mm]{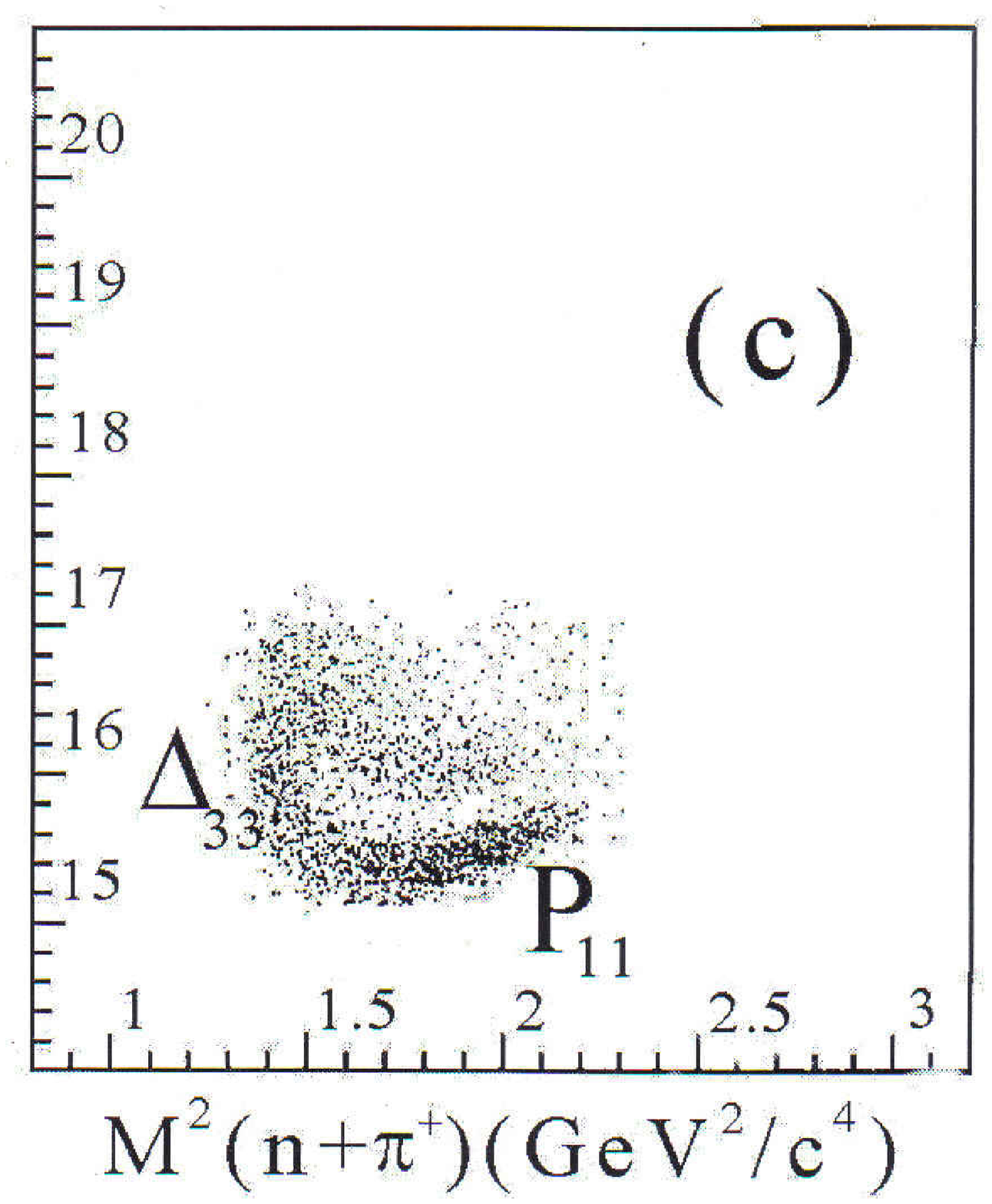}
\\
\parbox[t]{0.95\textwidth}{
\centering\caption{\small Dalitz plots for the $p(\alpha,
\alpha^\prime)n\pi^+$ reaction. (a) Simulated events of the
$\Delta$ resonance decay. (b) Simulated events of the Roper
resonance decay. (c) Experimental data. Here and in Fig. 5c, the
data of all SPES4 momentum settings, properly normalized, are
included.}}
\end{figure}

\begin{figure}[htb]
\includegraphics[width=0.31\textwidth,height=65mm]{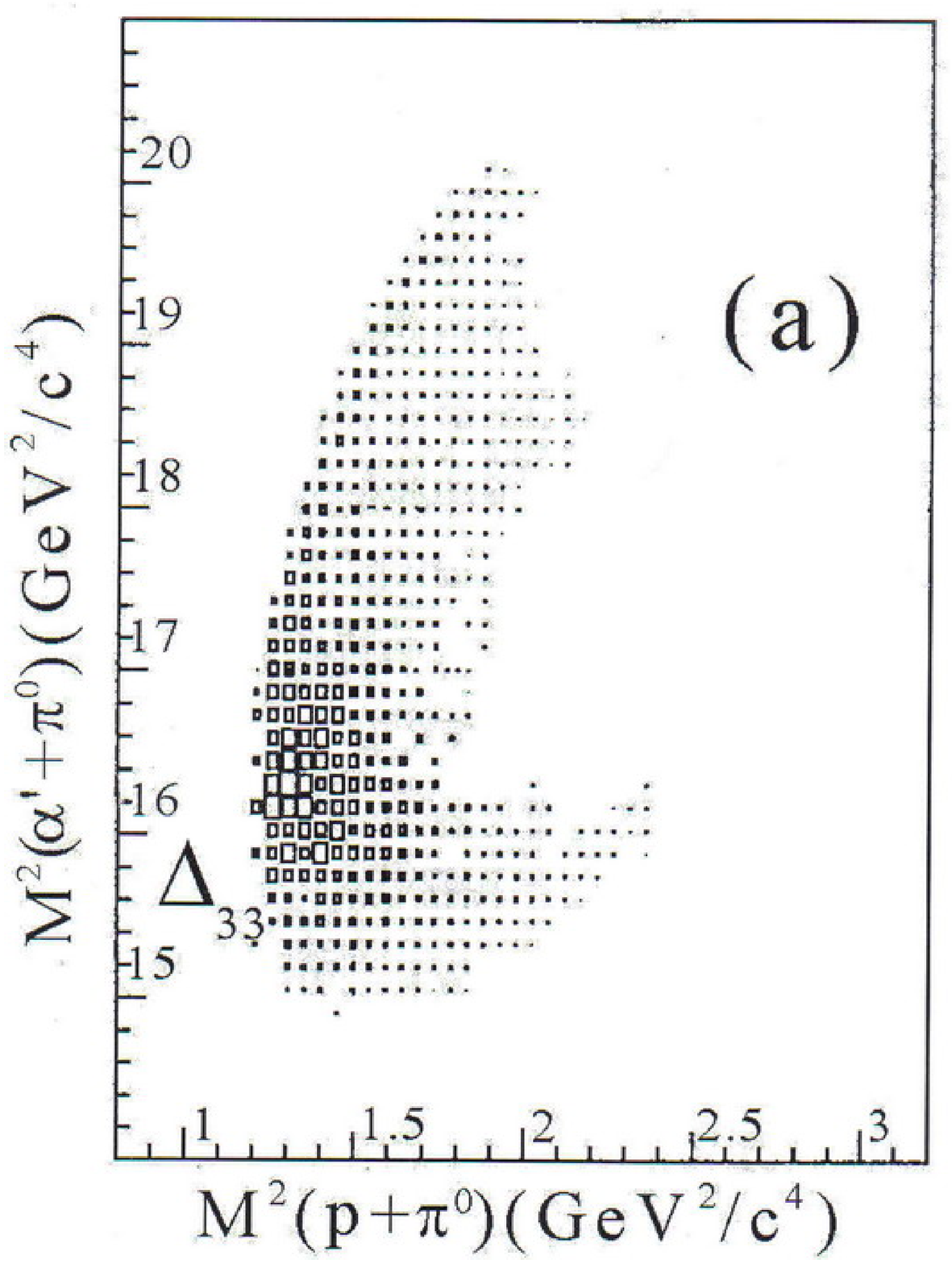}
\includegraphics[width=0.31\textwidth,height=65mm]{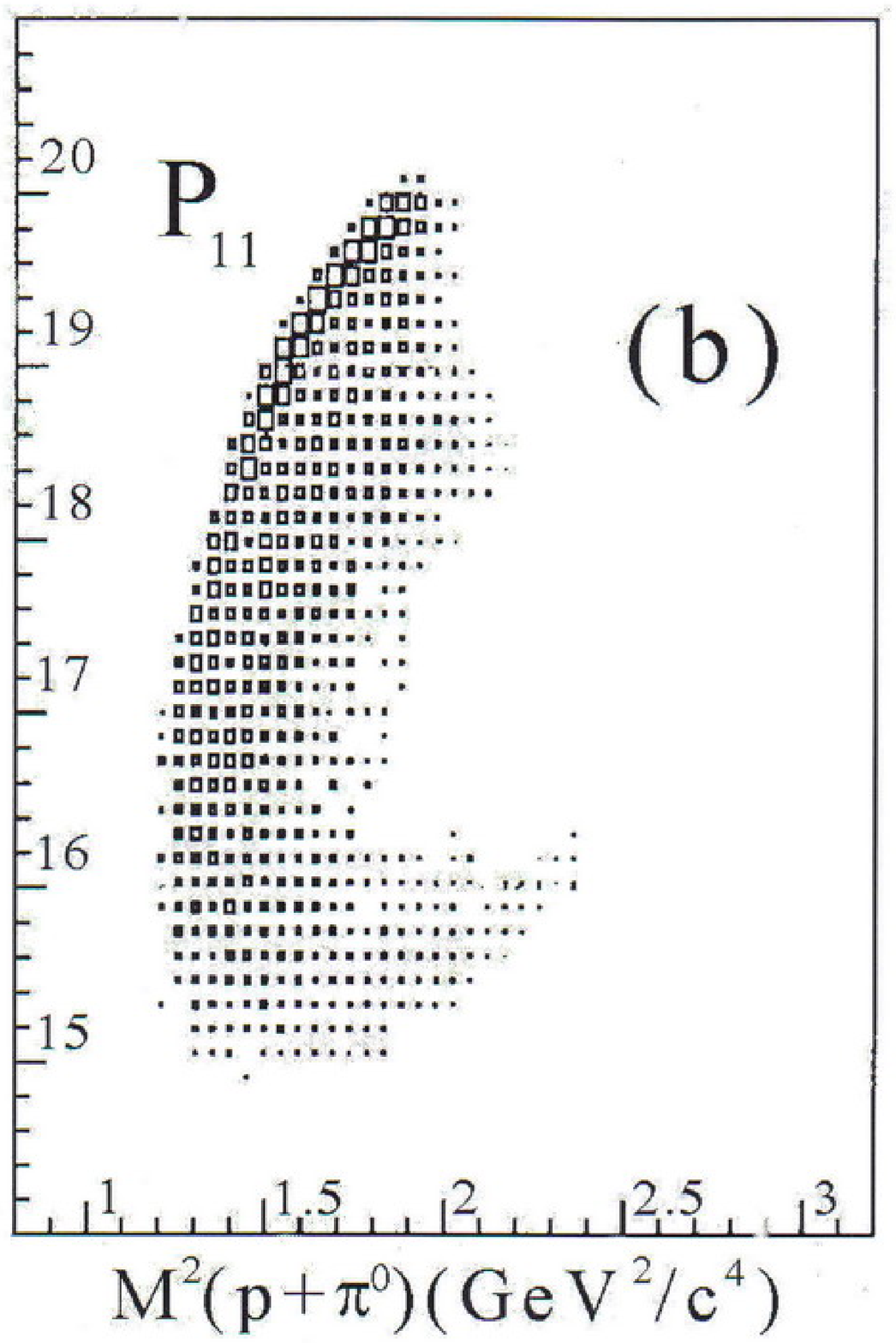}
\includegraphics[width=0.31\textwidth,height=65mm]{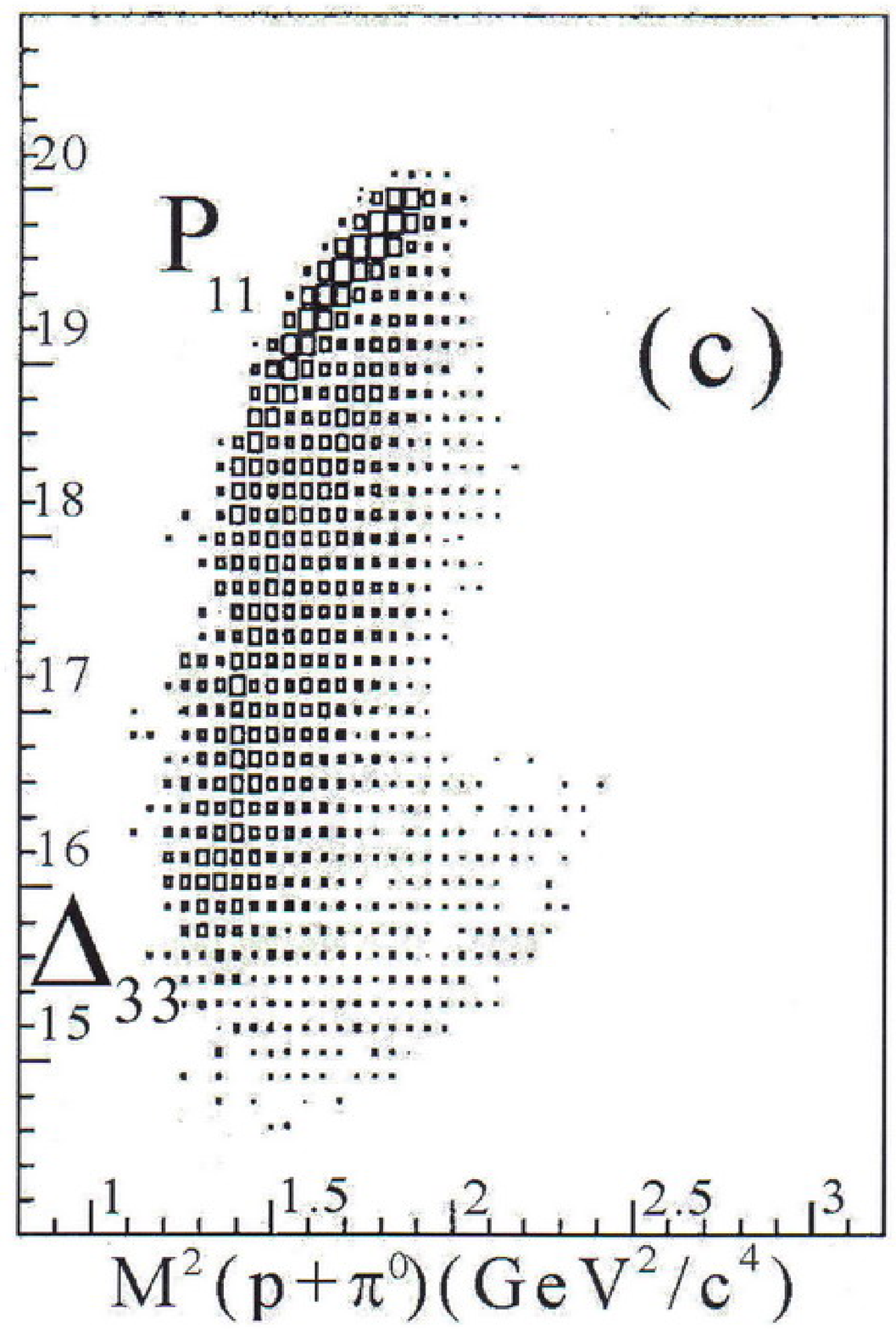}
\\
\parbox[t]{0.95\textwidth}{\centering
\centering\caption{\small Dalitz plots for the $p(\alpha,
\alpha^\prime)p\pi^0$ reaction. (a) Simulated events of the
$\Delta$ resonance decay. (b) Simulated events of the Roper
resonance decay. (c) Experimental data.}}
\end{figure}

The right panel of Fig. 3 presents the distributions of events as
a function of $M^2_{\rm miss}$ in the case of proton registration
by the FS. It is seen that for the SPES4 setting
$q_{\alpha^\prime}/Z$ = 3.35 GeV/$c$, corresponding to small
values of $|\omega|$, a peak at $M_{\rm miss}^2\simeq $ 0.02
(GeV/$c^2$)$^2$ (that is at $M_{\rm miss}\simeq M_{\pi}=$ 0.14
GeV/c$^2$) dominates in the spectrum. Evidently, this peak is due
to one-pion events mostly produced in the decay of the $\Delta$
resonance excited in the scattered $\alpha$ particle, as it was
discussed before. A slight tail at high masses in this spectrum is
presumably due to a small contribution of two-pion events from the
low-mass tail of the Roper resonance excited in the proton. The
width of the peak at $M_{\rm miss}^2\simeq$ 0.02 (GeV/$c^2$)$^2$
reflects the resolution of the reconstructed values of $M_{\rm
miss}^2$.

For the SPES4 momentum setting $q_{\alpha^\prime}/Z$ = 3.25
GeV/$c$, the contribution from two-pion events (at $M_{\rm miss}^2
\geq$ 0.09 (Gev/$c^2$)$^2$) is more prominent. In the interval of
0.04~$\leq$ $M_{\rm miss}^2$~$\leq$ 0.09~(GeV/$c^2$)$^2$, one-pion
and two-pion events are not resolved. As for the settings
$q_{\alpha^\prime}/Z$ = 3.15 and 3.06 GeV/$c$, the data show that
the two-pion production is an important channel of the inelastic
$p(\alpha,\alpha^\prime)pX$ reaction under study. While comparing
the numbers of the registered two-pion and one-pion events it
should be kept in mind that when we register protons or $\pi
^+$-pions we select different isospin projections of the studied
reaction. Also, the acceptances for detection of one-pion and
two-pion events in the considered cases are significantly
different. By imposing cuts on the values of $M_{\rm miss}$ (or
$M^2_{\rm miss}$) we can select
only one-pion or only two-pion events.\\

\begin{figure}[h]
\centering\epsfig{file=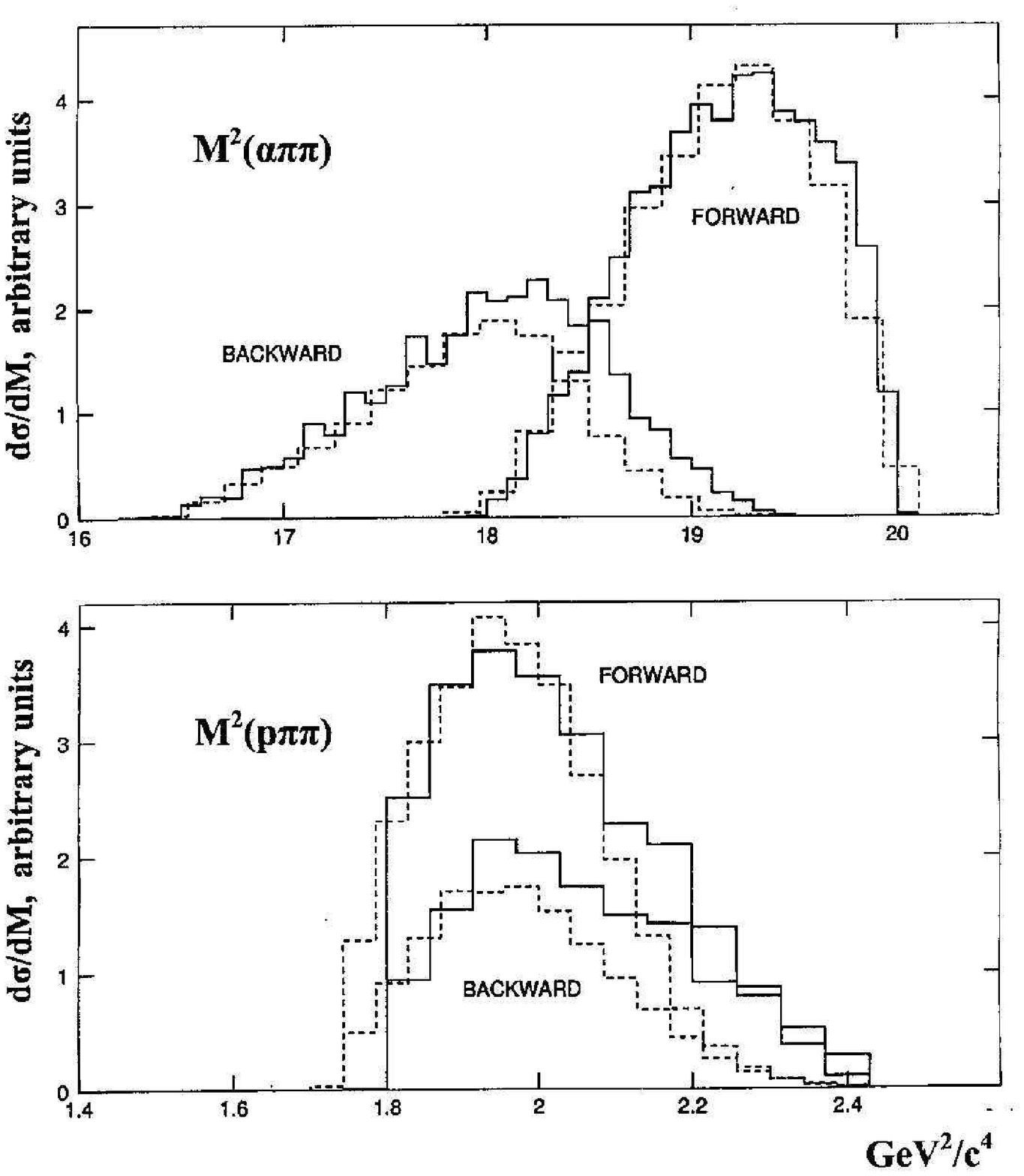,width=0.65\textwidth,height=80mm}
\caption{\small Comparison of the simulated invariant-mass-squared
$M^2(\alpha^\prime \pi\pi)$
 and $M^2(p\pi\pi)$ distributions (dashed line) with the distribution
 obtained from the experimental data (solid line) for the
 $p(\alpha,\alpha^\prime)p\pi\pi$ reaction for the forward and backward emitted
 protons in the $N^*$ centre-of-mass system at $q_{\alpha^\prime}/Z$ = 3.06~GeV/$c$.
 The MC simulations are performed assuming Roper excitation in the target.}
\end{figure}
\vspace{20mm}
\noindent {\bf 3.2 One-pion production}\\
\par
\noindent Figures 4 and 5 present the Dalitz plots of one-pion
events for the cases when pions (Fig. 4) and when protons (Fig. 5)
are registered with the FS. Comparing the Dalitz plots of the
experimental data (Fig. 4c and Fig. 5c) with the Dalitz plots of
the simulated events (Figs. 4a,b and Figs. 5a,b) we see that the
events in the spots where the experimental points are concentrated
in the Dalitz plots correspond to the simulated events from the
decay of the $\Delta$ and Roper resonances. Moreover, the
positions of these spots in the Dalitz plots tell us (in agreement
with theory) that the Roper resonance is excited in the target
proton, whereas the $\Delta$ resonance is excited in the
projectile $\alpha$ particle. In particular, the positions of the
maxima in the $M^2(N{\pi})$ spectra (as follows from Figs. 4c and
5c) for the Roper events are approximately the same in the cases
when pions and when protons are registered, as it should be for
the Roper resonance excitation in the target proton. On the other
hand, the positions of the maxima in the $M^2(\alpha^\prime \pi)$
spectra in these two cases are very different, which proves that
the Roper resonance is excited not in the projectile $\alpha$
particle (but it is excited in the target proton). Some depletion
of events in Fig. 5c in the region of $M^2(p\pi^0)$
$\approx$~1.3~$\div$~1.4~(GeV/$c^2$)$^2$ and $M^2(\alpha^\prime
\pi^0)$ $\approx$ 17 $\div$~18~(GeV/$c^2$)$^2$ can be interpreted
as an indication of a destructive interference between the
processes of one-pion production through excitation and decay of
the Roper and delta resonances. Note that according to theoretical
considerations of Hirenzaki {\it et al.} [10] the interference
between these processes gives a
negative contribution to the cross section.\\
\vspace*{10mm}
{\bf 3.3 Two-pion production: excitation of the Roper resonance}\\
\noindent Now we turn to the channel of two-pion production in the
case when protons are registered with the FS. It is evident that
the two-pion events are not from the decay of a $\Delta$
resonance, for which the branching for the two-pion decay mode is
very small. We can assume that the detected two-pion events are
due to excitation and decay of the Roper resonance in the target
proton. In order to check this conjecture, we have simulated the
spectra of the invariant squared masses
$M^2$($\alpha^\prime\pi\pi$) and $M^2(p\pi\pi$) for the
$p(\alpha,\alpha^\prime)p\pi\pi$ reaction and compared them with
the experimental data. The simulation calculations of these
spectra, as well as of the Dalitz plots discussed previously, were
performed with the phase space for the considered reactions
including the Roper and $\Delta$ resonances described by the
modified BW distribution with the mass-dependent resonance widths
according to Eqs. (9) and (11) of [15]. (We assumed the mass
dependence of the Roper-resonance width for two-pion decay to be
the same as that for one-pion decay.) The $\alpha$ form factor,
calculated using the parameterization of [3], and the SPES4-$\pi$
acceptance were also taken into account. The resonance masses and
widths of the Roper and $\Delta$ resonances were taken from a PDG
review [14]. To exclude a possible contribution of one-pion events
to the considered experimental spectra, only the events with
$M_{\rm miss}^2 \geq$ 0.09 (GeV/$c^2$)$^2$ were used. A similar
cut was also imposed on the simulated spectra. The shapes of the
simulated spectra are in satisfactory agreement with those of the
data, as it is demonstrated in Fig. 6 for the SPES4 momentum
setting $q_{\alpha^\prime}/Z$ = 3.06 GeV/$c$. Similar results were
also obtained for the setting $q_{\alpha^\prime}/Z$ = 3.15
GeV/$c$. Note that no fitting parameters were used in these
calculations, the Roper resonance parameters, as it was  said,
being taken from [14]: $M_R$ = 1440, $\Gamma_R$ = 350 MeV/$c^2$.

In principal, two pions might be produced in $\alpha p$ scattering
via intermediate-state excitation of the Roper resonance in the
projectile $\alpha$ particle, or double $\Delta$ excitations,
either both $\Delta$'s in the $\alpha$ particle, or one $\Delta$
in the target proton and one $\Delta$ in the $\alpha$ particle. We
have simulated such events. The shapes of the simulated spectra
for these reaction channels do not agree with those of the
experimental data (see [12]). We remind that according to theory
[10] the contributions from these channels are rather small, and
they may be neglected.
\begin{figure}[htb]
\centering \epsfig{file=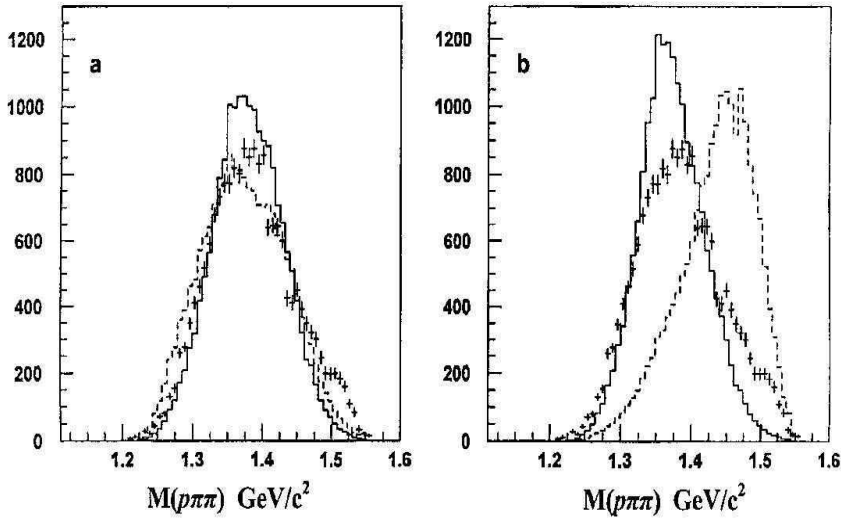,width=0.85\textwidth,height=80mm} {
\caption{\small Comparison of the simulated invariant-mass
distributions
 $M(p\pi\pi)$ (solid, dashed and dotted lines) with the experimental one (crosses)
 obtained from the data of the 3.06, 3.15 and 3.25~GeV/$c$ SPES4 momentum settings.
  Fig. 7a: dashed line -- phase-space
calculations, solid line -- Roper excitation with $M_R$ =  1440,
$\Gamma _R$ = 350~MeV/$c^2$ [14]. Fig. 7b: solid line -- Roper
excitation with $M_R$ = 1390, $\Gamma _R$ = 190~MeV/$c^2$ [3],
dotted line -- Roper excitation with $M_R$ = 1485, $\Gamma _R$ =
284~MeV/$c^2$ [16], dashed line -- $N(1520)D_{13}$ excitation with
$M_D$ = 1520, $\Gamma _D$ = 120~MeV/$c^2$ [14].
 The simulated spectra are normalized to the experimental one.}}
\end{figure}
Figure 7 presents a comparison of the simulated spectra of the
invariant mass $M(p\pi\pi$) with the corresponding experimental
spectrum obtained from the properly combined data of the SPES4
momentum settings $q_{\alpha^\prime}$/Z = 3.25,
 3.15 and  3.06~GeV/$c$. In these simulations, several BW Roper-resonance parameters
were used: from [14], [3] and [16]. One can see (Fig. 7a) that the
simulated spectrum of $M$($p\pi\pi$) is in reasonable agreement
with the experimental data when the standard Roper parameters are
assumed ($M_R$ = 1440, $\Gamma _R$ = 350~MeV/$c^2$ [14]). (Due to
influence of the $\alpha$ form factor the maxima in the simulated
$M(p\pi\pi)$ distributions are shifted to the masses smaller than
the values of the resonance masses used in the simulations.) The
results of the simulation with the Roper parameters from [3]
($M_R$ = 1390, $\Gamma _R$ = 190~MeV/$c^2$) are in somewhat worse
agreement with the data (see Fig.~7b). However, in view of not
sufficient precision of the data and due to some uncertainties in
the performed analysis, in particular due to an uncertainty in the
possible contribution of the higher-mass $N(1520)D_{13}$
resonance, our data analysis does not allow us to give preference
to one of these considered sets of the Roper parameters. As for
the $M$($p\pi\pi$) distribution simulated with the Roper
parameters from [16] ($M_R$ = 1485, $\Gamma _R$ = 284~MeV/$c^2$),
it is in noticeable disagreement with our data (see Fig. 7b).
According to a very recent partial-wave analysis of Sarantsev {\it
et al.} [17], the BW Roper-resonance parameters are: $M_R$ = 1436
$\pm$ 15, $\Gamma _R$ = 335 $\pm$ 40~MeV/$c^2$. On the other hand,
the new data of the BES collaboration on the $J/\psi$ decay [18]
and of the CELSIUM-WASA collaboration on the pion production in
$pp$ collisions [19] are in favour of smaller values of the Roper
mass and width: $M_R \simeq$ 1360, $\Gamma _R \simeq$
150~MeV/$c^2$.

We have also performed a simulation under an assumption that
two-pion events are produced via excitation and decay only of the
$N(1520)D_{13}$
 resonance. In this case, the results of the simulations are
in drastic disagreement with the data (Fig. 7b). At the same time,
it is seen that a small admixture of events from this resonance to
events from the Roper decay is possible. Adding to the simulated
spectrum $M$($p\pi\pi$) a small contribution of events from the
decay of the $N(1520)D_{13}$ resonance can improve the agreement
of the simulated spectrum with the data in the region of high
masses ($M(p\pi\pi)\simeq$ 1.5~GeV/$c^2$). According to our
estimation, the contribution of events from the $N(1520)D_{13} \to
p\pi\pi$ decay in the analyzed data may be about 10~$\div$~20$\%$.

Thus, we see that our data are consistent with the scenario that
two-pion events are produced mostly via excitation in the target
proton of the Roper resonance (with the mass of about
1390~$\div$~1440~MeV/$c^2$), which decays to a proton and two
pions. It should be admitted however that the shape of the
simulated $M$($p\pi\pi$) spectrum is also consistent with the data
for the case of non-resonant two-pion production (see Fig. 7a),
the difference between the shapes of the $M$($p\pi\pi$)
distributions for the non-resonant case and the resonant one (with
the Roper parameters from PDG) being relatively small. This may be
explained by the fact that the width of the Roper resonance is
large (as in PDG) and its propagator exerts little influence on
the shape of the simulated $M$($p\pi\pi$) spectrum. An estimate of
the non-resonant contribution has been made by Alvarez-Ruso {\it
et al.} [20] for the case of inelastic $pp$ scattering at 1~GeV.
It was shown that the non-resonant contribution should be about
two orders smaller than the resonant one. The same should be also
for $\alpha p$ scattering. Therefore, the non-resonant
contribution may be neglected. Taking this statement for granted,
and taking into account our previous considerations, we conclude
that the $p(\alpha,\alpha^\prime)p\pi\pi$ reaction (at an energy
of $\sim$~1~GeV/nucleon) proceeds mainly through the intermediate
state which is the Roper resonance excited in the target proton.
Due to the isoscalar nature of the $\alpha$ particle, the Roper
resonance, as has been already mentioned, may be excited in this
reaction via an exchange between the projectile $\alpha$ particle
and the target proton of a $\sigma$ meson, which is a coupled pion
pair in the
isospin $I$ = 0, $S$-wave state (Fig. 2c).\\

\noindent
{\bf 3.4 Two-pion production: decay of the Roper resonance}\\

\noindent In $\pi N$ scattering (see [14]), the two-pion decay of
the Roper resonance occurs mainly either as simultaneous emission
of two pions in the $I$ = 0 isospin, $S$-wave state, $N^* \to N
(\pi\pi)^{I=0}_{S-wave}$, or as sequential decay through the
$\Delta$ resonance, $N^* \to \Delta\pi \to N\pi\pi$, with
branching ratios of $\sim$~10$\%$ and $\sim$~30$\%$, respectively.
Manley {\it et al.} [4, 5] performing a partial-wave analysis of
the $\pi N \to N\pi~\&~N\pi\pi$ scattering data introduced a
$\sigma$ meson (or $\epsilon$ in the notation of [21]) as an
$S$-wave isoscalar $\pi\pi$ interaction.
 In the present analysis, we follow
Manley's approach to the two-pion decay of the Roper resonance and
also consider two possible channels of the Roper decay, one
through the $\Delta$ resonance and another one through the
$\sigma$ meson (Fig. 8).\\
\begin{figure}[htp]
\centering \epsfig{file=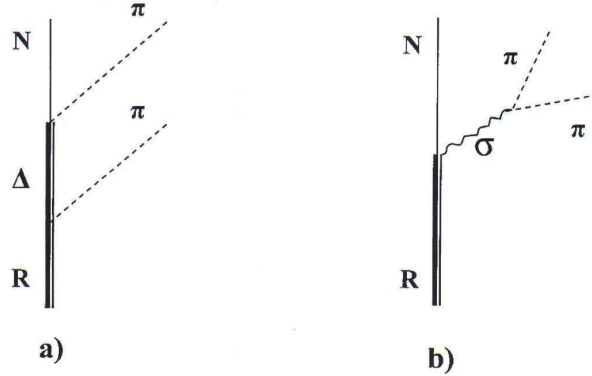,width=0.58\textwidth}
\caption{\small Diagrams for the Roper-resonance decay with
emission of two pions in Manley's approach according to [21]. (a)
Decay through the intermediate $\Delta$ state, $N^*\to\Delta\pi\to
N\pi\pi$. (b) Decay through the intermediate $\sigma$-meson state,
$N^*\to N\sigma\to N\pi\pi$.}
\end{figure}
As follows from theory [21], the shape of the spectra  of the
invariant mass $M$($\pi\pi$) of the pions emitted in the
Roper-resonance decay is essentially different for these two
channels. Therefore, a comparison of our experimental data with
theoretical predictions can be used to find out which process is
more important for decay of the Roper resonance excited in $\alpha
p$ inelastic scattering. In $\pi N$ scattering, according to [14],
a sequential $\Delta$ decay (Fig. 8a) is dominant. On the other
hand, as Morsch and Zupranski discussed [3], the breathing mode of
the nucleon is strongly excited in $\alpha p$ scattering, and a
different decay pattern (dominated by that shown in Fig. 8b) is
expected.

In Fig. 9, the simulated $M^2$($\pi\pi$) spectra are compared with
the experimental data for the SPES4 momentum settings
$q_{\alpha^\prime}/Z$ = 3.06 and 3.15 GeV/$c$, which have high
acceptance for events of the $p(\alpha,\alpha^\prime)p\pi\pi$
reaction.
 The experimental
spectra $M^2$($\pi\pi$) are obtained from the missing-mass-squared
$M_{\rm miss}^2$ spectra shown in Fig. 3 (right panel) by
subtracting the $M^2 (\pi)$ contributions of the
one-pion-production channels, the $M^2 (\pi)$ spectra being
parameterized by Gaussians (see dashed curves in Fig. 3). In these
simulations, the Roper, $\sigma$ and $\Delta$
 BW shapes, the $\alpha$ form factor, and the
SPES4-$\pi$ acceptance were taken into account. Further, the
simulated spectra were smeared to take into account the
experimental resolution of $M^2 (\pi\pi$), which was estimated
from the width of the $M^2 (\pi$) spectra. We have checked that an
uncertainty in the $\alpha$ form factor used affects the shape of
the simulated spectra only insignificantly. The following
parameters for the $\Delta$ resonance and $\sigma$ meson were
used: $M_{\Delta}$ = 1232, $\Gamma_{\Delta}$= 120~MeV/$c^2$ [14],
and $M_{\sigma}$ = 600, $\Gamma_{\sigma}$ =
 600~MeV/$c^2$ [22]. It should be noted that in the case of decay through the
intermediate $\sigma$ meson, the specific parameters of this meson
exert practically no influence on the simulated spectra due to the
large value of $\Gamma_{\sigma}$. As for the channel of decay
through the intermediate $\Delta$-resonance state, the amplitude
of this process is strongly influenced by the following
kinematical factor (see [21]):
\begin{equation}
A({\vec q_{\pi_1}}, {\vec q_{\pi_2}})\sim({\vec
q_{\pi_1}}\cdot{\vec q_{\pi_2}}),
\end{equation}
where ${\vec q_{\pi_1}}$ and ${\vec q_{\pi_2}}$ are the pion
momenta in the $N^*$ centre-of-mass system. We have not included
small spin-dependent terms in Eq. (1). According to [23, 24], the
contribution of these spin-dependent terms is about 16 times
smaller than that of the $\vec q_{\pi_1}\cdot\vec q_{\pi_2}$ term,
and to a good approximation it can be neglected. As a result of
the factor $A({\vec q_{\pi_1}}, {\vec q_{\pi_2}})$, the simulated
spectra of $M^2 (\pi\pi$) have two maxima, one near the minimum
values of $M$($\pi\pi$) (at $M(\pi\pi)$ close to 0.3~GeV/c$^2$)
and another one at larger values of $M$($\pi\pi$) (close to
0.45~GeV/c$^2$). The first peak corresponds to the events when
both emitted pions fly in the $N^*$ centre-of-mass system with
similar momenta in the same direction, while the second peak
corresponds to the events when the pions are emitted in opposite
directions.

\begin{figure}[h]
\centering \epsfig{file=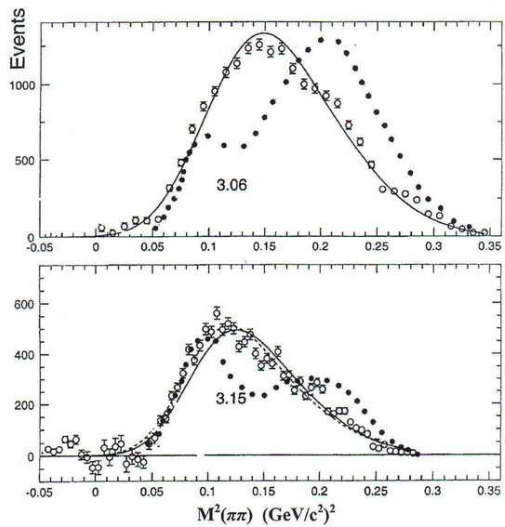,width=0.5\textwidth}
\caption{\small Invariant-mass-squared $M^2(\pi\pi)$ distributions
 for the
$p(\alpha,\alpha^\prime)p\pi\pi$ reaction. Open points --
experimental data. Solid curves -- results of the MC simulations
assuming the $N^*\to p\sigma \to p\pi\pi$ decay. Dotted curves --
results of the MC simulations assuming the $N^* \to \Delta\pi \to
p\pi\pi$ decay. Dashed line in the lower plot -- result of the MC
simulation assuming the $N^*\to p\sigma \to p\pi\pi$ decay with a
small admixture of events from the $N^*\to \Delta \pi\to p\pi\pi$
decay.}
\end{figure}
\begin{figure}[htb]
\centering \epsfig{file=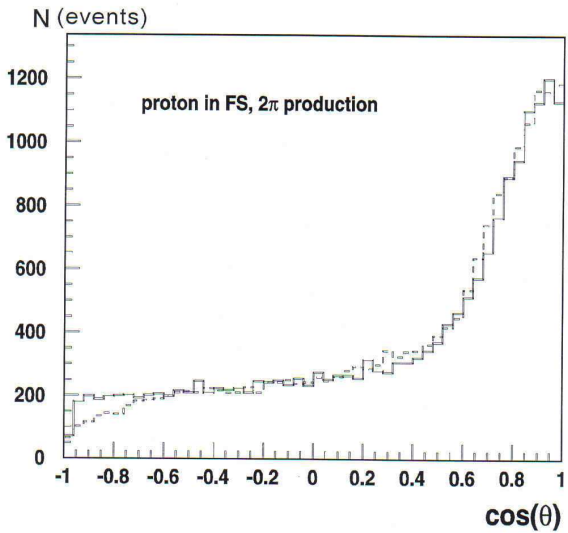,width=0.55\textwidth}
\caption{\small Angular distribution of the emitted protons in the
$N^*$ centre-of-mass system, not corrected for the SPES4-$\pi$
acceptance. Angle $\theta$ is the angle between the proton
momentum and the momentum transfer $\vec q_{\alpha}$ -- $\vec
q_{\alpha^\prime}$ in the rest frame of $N^*$. The solid line
shows the experimental data, the dashed line is the normalized
Monte Carlo simulation, made by assuming isotropic $N^*$ decay.
(The simulated spectrum, as well as the experimental one, is
distorted by the SPES4-$\pi$ acceptance.)}
\end{figure}

As one can see in Fig. 9, the shapes of the $M^2 (\pi\pi$) spectra
simulated for the channel of the Roper resonance decay through the
$\Delta$ resonance are in evident disagreement with the
experimental data for both SPES4 momentum settings. As opposed to
this, the shape of the simulated spectrum $M^2 (\pi\pi$) assuming
the decay through the intermediate $\sigma$ meson is in perfect
agreement with the data for the SPES4 setting
$q_{\alpha^\prime}/Z$ = 3.06 GeV/$c$. A similar spectrum for the
SPES4 setting $q_{\alpha^\prime}/Z$ = 3.15 GeV/$c$ is also in
fairly good agreement with the data\footnote[1]{Better agreement
with the data can be achieved in this case (see Fig. 9) if a small
admixture of events corresponding to the Roper decay through the
intermediate state of the $\Delta$ resonance is added to the
simulated spectrum.}. Thus, the $M^2 (\pi\pi$) spectra measured in
this experiment suggest that the Roper resonance excited in the
$p(\alpha,\alpha^\prime) p\pi\pi$ reaction at an energy of
$\sim$~1~GeV/nucleon decays mainly as $N^*\to p\sigma \to
p\pi\pi$.

This conclusion is also supported by the extracted angular
distribution of the emitted protons in the $N^*$ centre-of-mass
system. The obtained distribution agrees with the isotropic decay
of $N^*$ (see Fig. 10), and therefore it agrees with the assumed
picture of decay of the Roper resonance (with the spin 1/2) to a
nucleon and a scalar meson.

Our conclusion that the two-pion decay of the Roper resonance
excited in $\alpha p$~scattering at 1~GeV/nucleon proceeds
predominantly through the $N^* \to p\sigma \to p\pi\pi$ channel is
very different from previous $\pi N$-scattering-analyses results
[14]. On the other hand, our result nicely correlates with recent
investigations of the two-pion production in $pp$
inelastic-scattering experiments at energies of
0.65$~\div~$1.45~GeV [25]. The authors of these studies come to
the conclusion that the two-pion production in $pp$ scattering at
the considered energies proceeds mainly via excitation of the
Roper resonance, which decays predominantly through an
intermediate $\sigma$ meson. Sarantsev {\it et al.} [17] who
performed a combined partial-wave analysis of several
pion-production reactions also conclude that the channel of the
Roper decay through the $\sigma$ meson is rather important, the
contribution of this channel being about three times larger than
that given by PDG [14]. The importance of the $\sigma N$ channel
of the Roper decay was pointed out in several theoretical papers.
In particular, Dillig and Scott [26], considering the Roper
resonance in the constituent quark-gluon model, came to the
conclusion that the wave function of the Roper resonance contains
a very strong, $\simeq$ 50\%, component of a $\sigma$-meson field.
Consequently, the $\sigma N$ decay channel should dominate in the
two-pion decay of the Roper resonance. Similar statements that the
$\sigma N$ channel of the Roper decay is more important than the
$\pi \Delta$ channel, were also made by Kukulin {\it et al.} [27],
Krehl {\it et al.} [28], and some others.

As it was discussed by Morsch and Zupranski [3], the contribution
of the channel with the $\sigma$ meson can be different in
different two-pion-production reactions. Evidently, the properties
of the Roper resonance and the role of the $\sigma$
meson in pion-production reactions need further studies.\\

\noindent
{\bf 4 Conclusions}\\

\noindent The  one-pion and two-pion production $p(\alpha
,\alpha^\prime)X$ reaction has been studied in a semi-exclusive
experiment at the Saturne-II accelerator at an energy of $\simeq$
1~GeV/nucleon with the detection of the scattered $\alpha$
particle and the secondary pion or proton. The results of the
measurements are qualitatively compared with the simulated Dalitz
plots and invariant-mass spectra based on the predictions of the
Oset-Hernandez model using Manley's approach to the Roper decay.
 The obtained results show that the one-pion production in this reaction
 occurs via decay of the $\Delta$ resonance excited in the
 projectile $\alpha$ particle as well as decay of the Roper
 resonance excited in the target proton. The two-pion production
 occurs via decay of the Roper resonance excited in the target
 proton, the dominant channel of the Roper decay being $N^*\to
N\sigma\to N\pi\pi$. The obtained results are in favour of the
statement that the resonance excited in $\alpha p$ scattering at
the excitation energy
around 1440~MeV is the breathing-mode excitation of the nucleon.\\

\noindent {\bf Acknowledgements}\\

 We are indebted to all members of  the
SPES4-$\pi$ Collaboration:
 V.A.~Mylnikov, E.M.~Orischin,
B.V.~Razmyslovich, I.I.~Tkach, S.S.~Volkov, A.A.~Zhdanov
(Petersburg Nuclear Physics Institute, Gatchina, Russia),
W.~Augustyniak, P.~Zupranski (Soltan Institute for Nuclear{\small
Acknowled Physics, Warsaw, Poland), M.~Boivin, B.~Ramstein,
M.~Roy-Stephan (Lab. National Saturne, Saclay, France),
J.-L.~Boyard, L.~Farhi, Th.~Hennino, J.-C.~Jourdain, R.~Kunne
(Institut de Physique Nuclaire, Orsay, France), L.V.~Malinina,
E.A.~Strokovsky (Joint Institute for Nuclear Research, Dubna,
Russia), H.-P.~Morsch (Institut f\"ur
Kernphysik, Forschungzentrum J\"ulich, J\"ulich, Germany)\\

We also thank the Saturne-II accelerator staff for their help and
support during all phases of the program. The authors are
specially grateful to Prof.~A.A.~Vorobyov for his support and
valuable discussions. One of us (ANP) also acknowledges
Prof.~V.I.~Kukulin for a fruitful discussion.\\
\vspace*{10mm}
\noindent {\bf References}\\

\noindent {\small \hspace*{2mm}1. L.D. Roper, Phys. Rev. Lett.
{\bf 12}, 340 (1964).\\
\hspace*{2mm}2. H.P. Morsch {\it et al.}, Phys. Rev. Lett. {\bf
69}, 1336
(1992).\\
\hspace*{2mm}3. H.P. Morsch and P. Zupranski, Phys. Rev. C {\bf
61}, 024002
(1999).\\
\hspace*{2mm}4. D.M. Manley, R.A. Arndt, Y. Goradia and V.L.
Teplitz, Phys. Rev. D
{\bf 30}, 904 (1984).\\
\hspace*{2mm}5. D.M. Manley and E.M. Saleski, Phys. Rev. D {\bf
45}, 4002
(1992).\\
\hspace*{2mm}6. G. H\"ohler and  A. Schulte, $\pi N$ Newsletter
{\bf 7}, 94
(1992).\\
\hspace*{2mm}7. R.A. Arndt, W.J. Briscoe, I.I. Strakovsky, R.L.
Workman and M.
Pavan,\\
\hspace*{5mm}Phys. Rev. C {\bf 69}, 035213 (2004).\\
\hspace*{2mm}8. H.P. Morsch and P. Zupranski, Phys. Rev. C {\bf
71}, 065203
(2005).\\
\hspace*{2mm}9. H.P. Morsch, W. Spang and P. Decowski, Phys. Rev.
C {\bf 67},
064001 (2003).\\
10. S. Hirenzaki, P. Fern\'andez de C\'ordoba and E. Oset, Phys.
Rev.C {\bf 53}, 277 (1996).\\
11. P. Fern\'andez de C\'ordoba {\it et al.}, Nucl. Phys. A {\bf
586}, 586 (1995).\\
12. G.D. Alkhazov {\it et al.}, Phys. Rev. C {\bf 78}, 025205 (2008).\\
13. G.D. Alkhazov {\it et al.}, NIM  A {\bf 551}, 290 (2005).\\
14. {\it Review of Particle Physics}, J. Phys. G: Nucl. Part.
Phys. {\bf 33}, 1 (2006) and 2007\\
\hspace*{5mm} partial update for edition 2008
(URL:http://pdg.lbl.gov).\\
15. J.D. Jackson, Il Nuovo Cimento, {\bf XXXIV}, 1644 (1964).\\
16. R.A. Arndt, W.J. Briscoe, I.I. Strakovsky and R.L. Workman,\\
\hspace*{5mm}Phys. Rev. C {\bf 74}, 045205 (2006).\\
17. A.V. Sarantsev {\it et al.}, Phys. Lett. B {\bf 659}, 94
(2007).\\
18. M. Ablikim {\it et al.}, Phys. Rev. Lett., {\bf 97} 062001
(2006).\\
19. H. Clement {\it et al.}, arXiv:nucl-ex/0612015 v1 (2006).\\
20. L. Alvarez-Russo, E. Oset and E. Hern\'andez, Nucl. Phys. A
{\bf 633}, 519 (1998).\\
21. E. Hern\'andez, E. Oset and M. J. Vicente Vacas, Phys. Rev. C
{\bf 66}, 065201 (2002).\\
22. N.A. T\"ornquist, Sorynshiron and Kenken (Kyoto) {\bf 102},
E224 (2001).\\
23. Bo H\"oistad (CELSIUS-WASA Collaboration), Nucl. Phys. A {\bf
721}, 570c (2003).\\
24. J. P\"atzold, M. Bashkanov, R. Bilger and W. Brodowski, {\it
et al.},\\
\hspace*{5mm}Phys. Rev. C {\bf 67}, 052202(R) (2003).\\
25. T. Skorodko {\it et al.}, Eur. Phys. J. A {\bf 35}, 317(2008);\\
\hspace*{5mm}Progress in Particle and Nuclear Physics v.{\bf 61},
    issue 1, 168 (2008).\\
26. M. Dillig and M. Schott, Phys. Rev. C {\bf 75}, 067001 (2007).\\
27. V.I. Kukulin {\it et al.}, arXiv:0807.0192v1 [nucl-th](2008);\\
\hspace*{5mm}Annals of Physics {\bf 325}, 1173 (2010).\\
28. O. Krehl and J. Speth, Acta Physica Polonica B {\bf 29}, 2477
(1998); \\
\hspace*{5mm}O. Krehl, C. Hanhart, C. Krewald and  J. Speth, Phys.
Rev. C {\bf 62}, 025207 (2000).\\
}
\end{document}